\documentclass[twocolumn,tighten]{aastex62}
\pdfoutput=1 
\usepackage{amsmath,amstext}
\usepackage[T1]{fontenc}
\usepackage{apjfonts} 
\usepackage{xcolor}
\usepackage[figure,figure*]{hypcap}

\newcommand\aemulus{{\sc Aemulus}}
\newcommand\xihm{\xi_{\rm  hm}}
\newcommand\ximm{\xi_{\rm  mm}}
\newcommand\xihh{\xi_{\rm  hh}}
\newcommand*{\hmpc}{h^{-1}\textrm{Mpc}}

\newcommand*{\hgpc}{h^{-1}\textrm{Gpc}}
\newcommand{\hmsun}{h^{-1}\textrm{M}_{\odot}}

\newcommand{\df}{{\rm d}}

\shorttitle{The Aemulus Project IV: Emulating Halo Bias}
\shortauthors{McClintock, T., et al.}

\begin{document}

\title{The Aemulus Project IV: Emulating Halo Bias}

\author{Thomas McClintock}
\affiliation{Brookhaven National Laboratory, Bldg 510, Upton, NY 11973, USA}
\affiliation{Department of Physics, University of Arizona, Tuscon, AZ 85721, USA}

\author{Eduardo Rozo}
\affiliation{Department of Physics, University of Arizona, Tuscon, AZ 85721, USA}

\author{Arka Banerjee}
\affiliation{Kavli Institute for Particle Astrophysics and Cosmology and Department of Physics, Stanford University, Stanford, CA 94305, USA}
\affiliation{Department of Particle Physics and Astrophysics, SLAC National Accelerator Laboratory, Stanford, CA 94305, USA}

\author{Matthew R. Becker}
\affiliation{Kavli Institute for Particle Astrophysics and Cosmology and Department of Physics, Stanford University, Stanford, CA 94305, USA}
\affiliation{Department of Particle Physics and Astrophysics, SLAC National Accelerator Laboratory, Stanford, CA 94305, USA}
\affiliation{High Energy Physics Division, Argonne National Laboratory, Lemont, IL 60439, USA}

\author{Joseph DeRose}
\affiliation{Kavli Institute for Particle Astrophysics and Cosmology and Department of Physics, Stanford University, Stanford, CA 94305, USA}
\affiliation{Department of Particle Physics and Astrophysics, SLAC National Accelerator Laboratory, Stanford, CA 94305, USA}

\author{Sean McLaughlin}
\affiliation{Kavli Institute for Particle Astrophysics and Cosmology and Department of Physics, Stanford University, Stanford, CA 94305, USA}
\affiliation{Department of Particle Physics and Astrophysics, SLAC National Accelerator Laboratory, Stanford, CA 94305, USA}

\author{Jeremy L. Tinker}
\affiliation{Center for Cosmology and Particle Physics, Department of Physics, New York University, 4 Washington Place, New York, NY 10003, USA}
 
\author{Risa H. Wechsler}
\affiliation{Kavli Institute for Particle Astrophysics and Cosmology and Department of Physics, Stanford University, Stanford, CA 94305, USA}
\affiliation{Department of Particle Physics and Astrophysics, SLAC National Accelerator Laboratory, Stanford, CA 94305, USA}

\author{Zhongxu Zhai}
\affiliation{Center for Cosmology and Particle Physics, Department of Physics, New York University, 4 Washington Place, New York, NY 10003, USA}
\affiliation{IPAC, California Institute of Technology, Mail Code 314-6, 1200 E. California Blvd., Pasadena, CA 91125}

\begin{abstract}
\noindent Models of the spatial distribution of dark matter halos must achieve new levels of precision and accuracy in order to satisfy the requirements of upcoming experiments. In this work, we present a halo bias emulator for modeling the clustering of halos on large scales. It incorporates the cosmological dependence of the bias beyond the mapping of halo mass to peak height. The emulator makes substantial improvements in accuracy compared to the widely used \citet{Tinker2010} model. Halos in this work are defined using an overdensity criteria of 200 relative to the mean background density. Halo catalogs are produced for 40 $N$-body simulations as part of the \aemulus\ project at snapshots from $z=3$ to $z=0$. The emulator is trained over the mass range $6\times10^{12}-7\times10^{15}\ \hmsun$. Using an additional suite of 35 simulations, we determine that the precision of the emulator is redshift dependent, achieving sub-percent levels for a majority of the redshift range. Two additional simulation suites are used to test the ability of the emulator to extrapolate to higher and lower masses. Our high-resolution simulation suite is used to develop an extrapolation scheme in which the emulator asymptotes to the \citet{Tinker2010} model at low mass, achieving ${\sim}3\%$ accuracy down to $10^{11}\ \hmsun$. Finally, we present a method to propagate emulator modeling uncertainty into an error budget. Our emulator is made publicly available at \url{https://github.com/AemulusProject/bias_emulator}.
\end{abstract}



\section{Introduction}
\label{sec:introduction}

Analyzing the largest structures in the Universe, including galaxy groups and clusters, is a promising avenue for answering fundamental questions in cosmology. These objects form in the peaks of the large-scale structure (LSS), and their clustering properties contain a wealth of information about the contents of the Universe. In order to observe these structures in optical wavelengths one can use wide and deep photometric and spectroscopic surveys such as the Dark Energy Survey \citep[DES]{DESY1_KP}, the Subaru Hyper Suprime-Cam \citep[HSC]{HSC2018_Overview}, the Kilo-Degree Survey \citep[KiDS]{KiDSVik450_2019}, as well as upcoming surveys including the Large Synoptic Survey Telescope\footnote{\url{https://www.lsst.org}} (LSST), Euclid \citep{Euclid2011_Euclid} and the Dark Energy Spectroscopic Instrument\footnote{\url{https://www.desi.lbl.gov}} (DESI). Among the numerous challenges facing these projects in analyzing the LSS is dealing with the fact that the objects they observe are biased tracers of the underlying structure.

Galaxy clusters, galaxy groups, and galaxies of all masses live inside dark matter halos, which are biased tracers of the matter density field \citep{Kaiser84}. Halos form in the peaks of this field, which are clustered differently compared to the field as a whole. Hence, the ratio of the clustering of halos to that of the matter density field is referred to as the halo bias. High-mass halos that host galaxy clusters and bright galaxies are much more biased (strongly clustered) compared to low-mass halos, which on average form in regions of lower density. Therefore, using biased tracers to understand the density field of the Universe requires a model for the level of bias of these tracers. Using a bias model and measures of the statistical distribution of halos through clustering, lensing, or velocity dispersion can enable calibration of halo properties and their connection to the LSS \citep{Hu2006_ClusterLikelihoods,Baxter2016,Farahi2016,Jimeno2017_RMCC,McClintock2018,Murata2019_HSCClusterLensing}.

The halo bias is constant at large scales, meaning halos cluster proportionally to linear theory $\delta_{\rm h}=b\delta$ \citep{ShethTormen99,CooraySheth02}. In this regime, the halo bias is referred to as the {\it linear} halo bias. Models of the linear halo bias are critical in cluster cosmology. For example, one can use weak gravitational lensing around clusters at large scales \citep{Hayashi08, Zu14, Melchior2017, McClintock2018}, or the clustering of galaxy clusters \citep{Baxter2016,Jimeno2017_RMCC} to measure cluster masses. The halo bias can also be used to model galaxy clustering and probe the galaxy--halo connection \citep{WechslerTinker2018_GalaxyHaloConnection}. In all of these analyses, the results are sensitive to the chosen model of the halo bias \citep{ShethTormen99,ShethMoTormen01,Tinker2010}. In cluster cosmology studies, the halo bias model contributes to the final error budget on cosmological parameters and derived scaling relations. For instance, \citet{Baxter2016} inferred the mass--richness relation of clusters using measurements of their clustering in SDSS. The analysis was systematics limited by the 6 percent uncertainty in the halo bias model of \citet{Tinker2010}. Eliminating this source of systematic uncertainty would have brought the total error budget on the mass calibration in \citet{Baxter2016} down from 18 percent to 11 percent, making it a competitive measurement of the mass--richness relation. Therefore, reducing the systematic uncertainty of the halo bias is necessary as statistical precision improves over time.

A promising route for constructing precise models of halo statistics is using numerical simulation. Cosmological $N$-body simulations are our best method to make accurate predictions of the structure of the Universe given a set of cosmological parameters. The distribution of particles and halos in simulations are used to calibrate accurate and precise models of the LSS \citep{PressSchechter74, Smith2003_halofit, Wechsler01, Wechsler2006_AssemblyBias, Takahashi2012_halofit, Heitmann2009_Coyote2, Mead2015_HaloModel, DarkQuest1_2018, Smith2018_NGenHalofit, DeRose2018}. For halos, these models can take the form of analytic functions \citep{ShethTormen99, Jenkins01, Tinker2008, Tinker2010} or more advanced predictive tools such as emulators \citep{Heitmann2016_MT1,DarkQuest1_2018,McClintock2018_HMF,Zhai2018,Euclid2018_Emulator}. Emulators are probabilistic interpolators that connect measurements from $N$-body simulations run with a specific set of cosmological parameters, allowing a user to predict that measurement at a new location in parameter space. Emulators are useful when analytic models are difficult or impossible to derive \citep[e.g.][]{Biswas2019_NuDEHalos}, and current emulators have achieved a level of precision in several statistical descriptions of the LSS that is sufficient to model 
much of the next generation of experiments \citep{Lawrence2017_MT2,McClintock2018_HMF}.

This work presents an emulator model for the linear halo bias that enhances the analytic model presented in \citet{Tinker2010}. The emulator is constructed from the simulations in the \aemulus\ project \citep{DeRose2018}, previously used for emulating the halo mass function \citep{McClintock2018_HMF} and galaxy correlation function \citep{Zhai2018}. In our emulator, precise measurements of the halo bias in the simulations are interpolated across cosmological parameter space, so that the halo bias at any point in parameter space within the simulation cloud can be obtained. Extrapolations beyond the limits of the masses in our simulation asymptote to the \citet{Tinker2010} bias model. In this work we compute the accuracy of our halo bias emulator as a function of mass and redshift between $10^{13}-7\times10^{15}\ \hmsun$ and $z\in[0,3]$. We also provide a tool to produce realizations of correlated noise. This can be used to propagate the modeling uncertainty into error budgets that use the emulator. The emulator is publicly available\footnote{\url{https://github.com/AemulusProject/bias_emulator}}.

\begin{figure}
	\begin{center}
	\includegraphics[width=\linewidth]{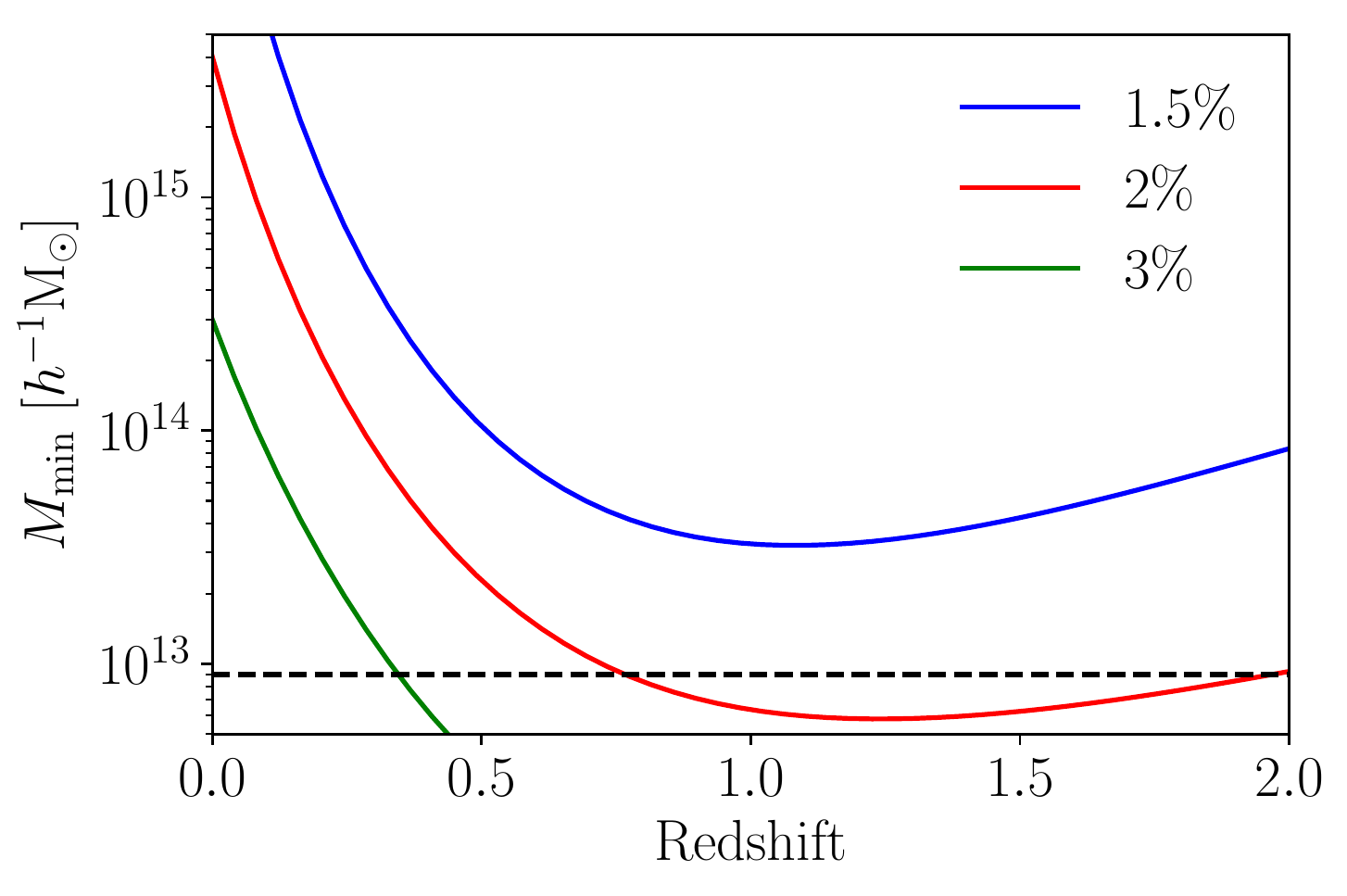}
	\caption{The minimum halo mass modeled by our emulator as a function of redshift at fixed precision in halo mass measurements. The applicability of our emulator depends on the level of precision achievable given some data. The black dashed line shows the minimum halo mass resolved in our training simulations used to construct the emulator. Our emulator can successfully be used without contributing to a final error budget if mass calibration is performed at or greater than the 2 percent level. A derivation of the curves appearing in this figure appears in \autoref{app:required_bias_emulator_accuracy}.
	\label{fig:bias_accuracy}}
	\end{center}
\end{figure}

The required accuracy of a halo bias model depends on the achievable precision in mass calibration in real data. \autoref{fig:bias_accuracy} shows the minimum halo mass for which our emulator can be used to model the halo bias while negligibly contributing to the final error budget in an analysis. The mass at a given redshift is set by the accuracy of our emulator (see \autoref{sec:building_an_accuracy_model}) as well as the precision to which one is able to calibrate halo masses. Three curves, corresponding to mass calibrations of 1.5, 2 and 3 percent are shown in \autoref{fig:bias_accuracy} as the blue, red and green lines, respectively. The black dashed line shows the minimum halo mass resolved in the simulations used to construct our emulator, as discussed in the following section. State-of-the-art cluster mass calibration analyses reach $\approx 5\%$ uncertainties, and are dominated by systematic errors.  Thus, the bias calibration presented here is sufficiently precise for current cluster samples, and will remain so for at least the next few years.  Whether our bias emulator will suffice for the LSST era will hinge on what the ultimate systematics floor of the data is.

In \autoref{sec:simulations} we detail the simulations and clustering measurements. We outline the training of the Gaussian processes that make up the emulator in \autoref{sec:constructing_the_emulator}, and validate its performance in \autoref{sec:emulator_accuracy}. \autoref{sec:building_an_accuracy_model} provides a model for the accuracy of the emulator that can be used to make random realizations of residuals about the model. \autoref{sec:using_the_emulator} contains a brief discussion of how to apply the emulator, in order to recover $\xihm$ or $\xihh$. \autoref{sec:comparison_to_other_simulations} tests the emulator against simulations run with larger volumes and higher resolution than those used for training and details the extrapolations beyond the mass limits of the training simulations. Finally, \autoref{sec:conclusions} summarizes the main results of this work. In \autoref{app:required_bias_emulator_accuracy} we derive the curves shown in \autoref{fig:bias_accuracy}, while in \autoref{app:bias_parameters_common_cosmos} and \autoref{app:hmf_parameters_common_cosmos} we supply halo bias and halo mass function emulator predictions for a selection of cosmological models. 

In all equations, $\ln$ refers to the natural log while $\log$ is the logarithm with base 10. Masses are defined by
\begin{equation}
	\label{eq:mass_definition}
	M_{\Delta} = \Delta\frac{4\pi}{3}R_{\Delta}^3 \Omega_m\rho_{\rm crit}\,,
\end{equation}
where $\rm \Delta=200$ is 200 times the background matter density throughout and $R_{\Delta}$ is the radius of the halo. Unless otherwise stated, distances are $\hmpc$ comoving and masses are $\hmsun$.

\section{Simulations}
\label{sec:simulations}

This work uses four sets of simulations produced for the \aemulus\ project. Two sets include the ``training simulations'' and the ``test simulations''. The other two sets are additional testing simulations run with more volume and higher resolution than the training and test suites. The training set consists of 40 simulations with different cosmological parameters used to construct and train our emulators. The test set consists of 35 additional simulations run at seven different locations in cosmological parameter space. In the test set, at each point in parameter space there are five realizations of each simulation seeded with different initial conditions. When combined, this reduces the sample variance by a factor of five. For all simulations, we saved snapshots at ten epochs between $z=3$ and $z=0$ at redshifts $z\in[3, 2, 1, 0.85, 0.7, 0.55, 0.4, 0.25, 0.1, 0]$. The testing and training simulations were run with $1400^3$ particles in cubic boxes of volume $(1.05\ \hgpc)^3$.

The simulations exist in a 7-dimensional cosmological parameter space. These parameters are the Hubble constant $H_0$, matter density fraction $\Omega_m$, baryon density fraction $\Omega_b$, dark energy equation of state $w$, primordial power spectrum index $n_s$, variance on 8 $\hmpc$ scales at $z=0$ parameterized by $\sigma_8$, and effective number of relativistic species in the primordial plasma $N_{\rm eff}$. The amplitude of the primordial power spectrum $A_s$ is also recorded.

The training simulations span the $3\sigma$ posterior distribution contours of the combination of four data sets. These are Planck+WMAP+BAO+SNIa \citep{PlanckXVI,WMAP9,Anderson2014,Suzuki2012}. The cosmologies of the testing simulations reside within the cloud of the training simulations, but do not overlap exactly. Details regarding the initial conditions of the simulations, choices concerning force softening and time stepping, and the convergence of various summary statistics measured using the halos and dark matter particles are discussed in \citet{DeRose2018}.

The larger and smaller volume simulations contain $2048^3$ particles, and were used to investigate resolution effects and test the ability of the emulator to extrapolate to higher and lower masses. One set of such simulations have volumes of $(400\ \hmpc)^3$ and are part of a separate high-resolution simulation suite.  The other set has volumes of $(3\ \hgpc)^3$, run with the same cosmological parameters as the test simulations. Neither set of simulations were used in the construction of the emulator.

Halos were identified with the \textsc{rockstar} halo finding algorithm \citep{Behroozi2013}, modified for our purposes \citep[see][for more information]{DeRose2018}. We only consider halos with at least 200 particles to avoid resolution effects. Particle masses $M_{\rm p}$ are given by  
\begin{equation}
    \label{eq:particle_mass}
    M_{\rm p} = \frac{V \rho_{\rm crit}}{N_{\rm p}}\frac{\Omega_m}{0.3}\,,
\end{equation}
where $V$ is the volume, $N_{\rm p}$ is the number of particles and $\rho_{\rm crit}$ is the critical density. The maximum halo mass probed by our training simulations was $7\times10^{15}\ \hmsun$, while the minimum halo mass was $6\times10^{12}\ \hmsun$. In \citet{McClintock2018_HMF} we found that the mass function depended on particle resolution, thus requiring a slight correction at low masses. Here, we made no correction to the halo bias because we found that particle resolution did not have a significant effect on the clustering.

We note that details of producing a halo catalog will affect measurements of the halo bias. For instance, using halos identified with the friends-of-friends (FOF) algorithm will yield halo populations with different mass functions and clustering properties than \textsc{rockstar} \citep{Jing1999_LagrangianBias,ShethTormen99,ShethMoTormen01,SeljakWarren04,Tinker2008,Pillepich2010_nongaussianICs,Reed2009_highzbias}. Additionally, subtle choices such as halo percolation can have effects at the 10\% level \citep{GarciaMar2019_Percolation}. We caution users of our emulator to be aware of these potential issues, and take care when using halo catalogs constructed using different definitions and algorithms.


\subsection{Halo Bias}
\label{sec:halo_bias}

We define the halo bias as the ratio of the halo--matter correlation function to the matter correlation function 
\begin{equation}
	\label{eq:halo_definition}
    b(r,M,z) = \frac{\xi_{\rm hm}(r,M,z)}{\xi_{\rm mm}(r,z)}\,.
\end{equation}
At linear (large) scales, the bias is scale independent, and depends most strongly on halo mass $M$. In this regime, our definition is equivalent to working with power spectra at small wavenumber $k$. We do not consider secondary bias dependencies \citep[concentration, accretion rate, etc.;][]{Wechsler2001_Thesis,Gao2005_AssemblyBias,Wechsler2006_AssemblyBias,Li2008_AssemblyBias,Mao2018_AssemblyBias,Contreras2019_AssemblyBias,Mansfield2019_AssemblyBias,Han2019_AssemblyBias}. 
In practice, we measure the linear halo bias for halos selected by mass as the ratio of the halo--matter correlation function $\xihm$ to the matter--matter correlation function $\ximm$ over a specific radial range given by
\begin{equation}
	\label{eq:halo_definition_xi}
    b(M,z) = \frac{\xi_{\rm hm}}{\xi_{\rm mm}}\bigg|_{\rm 10-40\ {\it h}^{-1}Mpc}\,.
\end{equation}
The exact scales for which this relation holds is cosmology and epoch-dependent. As discussed in \autoref{sec:clustering_measurements}, we verified that the above relationship is scale-independent within the redshift range of interest. The lower limit was chosen to avoid the 1-halo regime at all mass scales. We chose an upper limit of $40 \hmpc$ to accommodate our jackknife subregions and because above this scale the clustering becomes noise dominated. However, we checked that the ratio is scale-independent past 100 $\hmpc$.


\subsubsection{Clustering measurements}\label{sec:clustering_measurements}

We measure the clustering in our snapshots using tool Corrfunc\footnote{\url{https://corrfunc.readthedocs.io/en/master}} \citep{sinha_CORRFUNC}. The measurement spans 50 logarithmically spaced comoving radial bins between 0.1 and 80 $\hmpc$. The lower limit is five times the softening length of 0.02 $\hmpc$. Above 80 $\hmpc$ the clustering is noise dominated for most halo masses. We divided the total volume of $1050\ (\hmpc)^3$ into 1000 subregions, with subregions having $105\ \hmpc$ per side. We use the method of \citet{GarciaJK} to jackknife without using catalogs of random points. The clustering measurements will be discussed in greater detail in a future work focused on constructing correlation function emulators. The matter clustering and halo--matter clustering are converged to better than 1\% on all scales considered here \citep{DeRose2018}. We confirmed that the scale-dependent bias converged as well.

Halos were divided into mass bins using a sliding window function in the log that ensured sufficient statistics of the $\xihm$ signal for all mass bins. The highest mass bin was fixed to always have at least 500 halos, and mass bins descended from that threshold to the minimum mass of the particular simulation according to
\begin{equation}
	\label{eq:sliding_mass_window}
    \log M_{i-1} = \log M_{i} - 0.15 + 0.1\frac{\log M_{i} - \log M_0}{\log M_{\rm min} - \log M_{0}}\,
\end{equation}
where $\log M_{i}$ is the left edge of the $i$th mass bin, $\log M_{0}$ is the left edge of the highest mass bin, and $\log M_{\rm min}$ is the minimum halo mass of the simulation. The number of mass bins varied between snapshots, ranging from $\sim 4$ at $z=3$ to $\sim 25$ at $z=0$. There are 11 radial bins in the radial range we use to measure the halo bias. Therefore, jackknifing yields a numerically stable estimate of the covariance matrix for these scales.

\begin{figure*}
	\begin{center}
	\includegraphics[width=\linewidth]{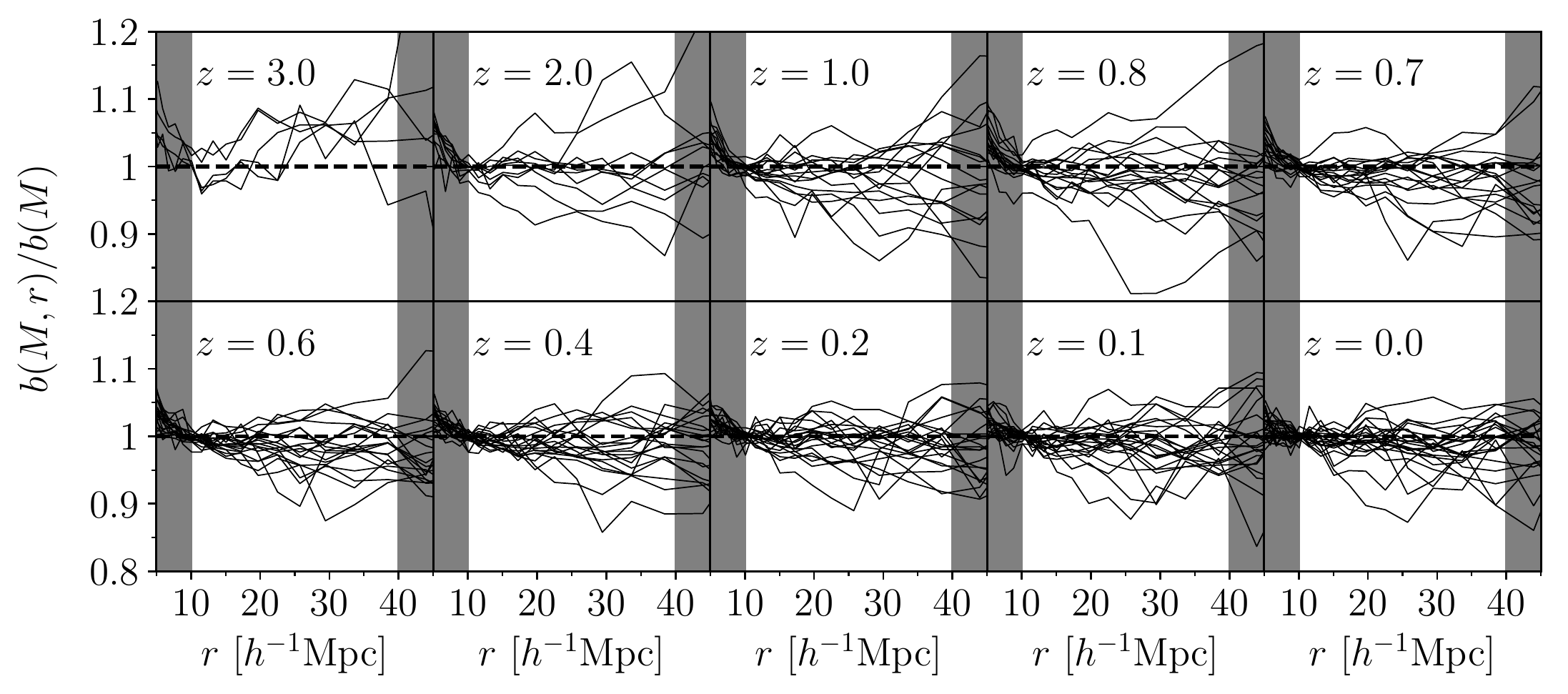}
	\caption{Ratio of the measured scale-dependent linear halo bias $b(M,r)$ to the constant bias fit from \autoref{eq:bias_likelihood} for all mass bins in all snapshots in one training simulation. The constant model was fit over the scales $10-40\ \hmpc$, and excluded scales are in gray. The lower limit avoids the 1-halo term, while larger scales are noisy and add little to the fit. Uncertainties on individual points (vertices of the lines) are not shown for clarity, but are of the order of the scatter. This means the halo bias is consistent with a constant at these scales.
		\label{fig:bias_ratio_figure}}
	\end{center}
\end{figure*}

Rather than propagating the uncertainty on both $\xihm$ and $\ximm$ forward into the bias, we jackknife the ratio of the two in order to estimate the covariance of the scale-dependent bias. The linear scale-independent bias is then determined by modeling the bias in \autoref{eq:halo_definition_xi} as a constant between 10 and 40 $\hmpc$. For a given mass bin in a snapshot of a simulation, the scale-independent bias $b(M)$ was found by maximizing the likelihood
\begin{equation}
	\label{eq:bias_likelihood}
    \ln\mathcal{L}_b \propto -\frac{1}{2}\Delta b^T(M,r)\ {\bf C}_r^{-1}\ \Delta b(M,r)\,. 
\end{equation}
In this equation, $\Delta b(M,r) = b(M,r) - b(M)$ is the difference between the bias at scale $r$ and our scale-independent model, and ${\bf C}_r$ is the covariance matrix of the scale-dependent bias over the radial scales we consider. By maximizing this probability distribution we obtained the best fit bias $b(M)$ for each mass bin.

We repeated this process leaving out each spatial jackknife subregion. This allowed us to compute the jackknife covariance matrix of the linear bias between mass bins ${\bf C}_b$. In the test simulations we computed the weighted mean of the five realizations for $b(M)$, and averaged the realizations of the covariance matrix ${\bf C}_b$. Since we combined five realizations of each testing simulation, we then divided ${\bf C}_b$ by a factor of five.

The ratio of $\xihm/\ximm$ to $b(M)$ for one simulation appears in \autoref{fig:bias_ratio_figure}. Uncertainties on the measurements have been removed for clarity. The $\chi^2$ of these fits are all acceptable, meaning the bias at these scales are well described by a constant model.


\subsection{Fitting the \aemulus\ simulations}
\label{sec:fitting_the_simulations}

The \citet{Tinker2010} halo bias model is commonly used in modern clustering analyses. That model provides a fitting function for the halo bias as a function of peak height $\nu(M,z)$. We use the same fitting function in our model given by
\begin{equation}
	\label{eq:tinker2010_fitting_function}
    b(\nu) = 1- A\frac{\nu^a}{\nu^a - \delta_c^a} + B\nu^b + C\nu^c\,,
\end{equation}
where $\nu(M,z) = \delta_c/\sigma(M,z)$ is the peak height of a halo of mass $M$ at redshift $z$, and $\delta_c=1.686$ is the critical density for collapse. Halos of mass $M$ are associated with a Lagrangian radius $R$ by $M = 4\pi \rho_mR^3/3$, meaning we can calculate the RMS mass variance of the linear density field associated with this length scale from
\begin{equation}
	\label{eq:sigma_definition}
    \sigma^2(R,z) = \int_0^{\infty}\frac{\df k}{k}\ \frac{k^3P(k,z)}{2\pi^2}|\hat{W}(kR)|^2\,,
\end{equation}
where $P(k,z)$ is the linear matter power spectrum and $\hat W$ is a real-space top-hat window function transformed into Fourier space. We use the \textsc{CLASS}\footnote{\url{http://class-code.net/}} Boltzmann code to compute the power spectrum \citep{CLASS1,Takahashi2012_halofit,Smith2003_halofit}. 
Unlike the approach of \citet{Tinker2010}, we incorporate non-universality explicitly by allowing the free parameters of the fitting function to vary with cosmological parameters. This is identical to the approach of \citet{McClintock2018_HMF}, where they allowed the fitting function parameters in the \citet{Tinker2008} mass function to vary with cosmological parameters.

We fit all mass bins in all snapshots in a given simulation simultaneously. Since the mass bins have a finite width, we model the bias in a single mass bin according to 
\begin{equation}
	\label{eq:bias_in_bin}
    \langle b \rangle = \frac{\int_{M_{\rm min}}^{M_{\rm max}} \df M\ b(M)\frac{\df n}{\df M}}{\int_{M_{\rm min}}^{M_{\rm max}} \df M\ \frac{\df n}{\df M}}\,,
\end{equation}
where we have integrated over the mass function $\df n/\df M$, using our emulator presented in \citet{McClintock2018_HMF}, as part of the \aemulus\ Project. For a given simulation, we found the posterior distribution of the parameters in \autoref{eq:tinker2010_fitting_function} by exploring the likelihood
\begin{equation}
	\label{eq:full_bias_likelihood}
    \ln\mathcal{L} \propto -\frac{1}{2}\sum_{i}^{N_z} \Delta {\bf b}_{i}^T {\bf C}_{b;\ i}^{-1} \Delta {\bf b}_{i} \,,
\end{equation}
where the sum runs over all snapshots and $\Delta {\bf b} = {\bf b}_i - \langle {\bf b} \rangle_i$ is a vector containing the difference between the scale-independent biases in each mass bin and the model in each bin given by \autoref{eq:bias_in_bin}.

Each of the free parameters in \autoref{eq:tinker2010_fitting_function}, $p\in(A,a,B,b,C,c)$ could, in principle, vary with redshift. To first order, this variation is well described as a linear function of the scale factor:
\begin{equation}
	\label{eq:parameter_scalefactor_dependence}
    p(z) = p_0 + (a-a_{\rm pivot})p_1\,,
\end{equation}
where $p_0$ is the value of the given parameter at $z=1$ while $p_1$ is the slope with scale factor $a = 1/(1+z)$. The pivot scale factor is $a_{\rm pivot}=0.5$, at redshift $z=1$. In practice, allowing all twelve free parameters to vary allows for too much flexibility. We allow $p\in(B_0, c_0, A_1, B_1)$ to vary and keep the rest held fixed to constant values. These constant values were found by maximizing the likelihood for an individual simulation allowing for more than the fiducial 4 parameters to vary. The specific constant values are not unique, since strong degeneracies exist between many of them if they are allowed to vary. For completeness they are presented in \autoref{tab:parameter_values}.

\begin{table}[ht]
	\centering
	\setlength{\tabcolsep}{.4em}
	\caption{Values of the fitting function parameters in \autoref{eq:tinker2010_fitting_function} and \autoref{eq:parameter_scalefactor_dependence} which are held fixed as a function of cosmological parameters. \label{tab:parameter_values}}
	\begin{tabular}{l|llllllll}
	    Parameter & $A_0$ & $a_0$ & $b_0$ & $C_0$ & $a_1$ & $b_1$ & $C_1$ & $c_1$\\ \hline
	    Value &  4.28 &  0.47 &  1.52 &  0.89 &  -0.56 &  -0.63 &  -0.60 &  -1.85\\
	\end{tabular}
\end{table}

As was the case when constructing the halo mass function emulator in \citet{McClintock2018_HMF}, in a given simulation the free parameters of the fitting function are correlated with each other. This is problematic because the Gaussian processes that interpolate these parameters as a function of cosmological parameters cannot account for these correlations. To mitigate this, we calculated the covariance between the fitting function parameters in the central most simulation, \aemulus\ index 34, and found the rotation matrix from its eigenvectors that diagonalized this covariance. We applied this rotation matrix to the chains from all simulations to obtain a new set of parameters that were orthogonal to each other, which we label $p'\in(B_0', c_0', A_1', B_1')$. We checked that this rotation matrix exhibited only a weak dependence on cosmology, and that our fiducial choice of which simulation to use to compute the rotation matrix did not affect the final results. Applying the inverse of this rotation matrix recovers parameters of the fitting function.

At this point we obtained a set of four independent parameters with error estimates describing the halo bias for each of the 40 training simulations. With these parameters we were able to construct our emulator.


\section{Constructing the emulator}
\label{sec:constructing_the_emulator}

In order to interpolate the fitting function parameters as a function of cosmological parameters, we use a set of four Gaussian processes, one for each of $p'\in (B_0', c_0', A_1', B_1')$. Gaussian processes are tools to perform probabilistic regression given a set of training data. \citet{RasmussenWilliamsGPs} provides a complete discussion on the topic. Here we provide a brief description of Gaussian process regression. We use the Gaussian process implementation provided in the \textsc{Python} package \texttt{george}\footnote{\url{http://george.readthedocs.io/en/latest}} \citep{Ambikasaran2014}.


\subsection{Gaussian process basics}
\label{sec:gp_basics}

Consider the goal of predicting the value of a function $y^*(x^*)$ at some input location $x^*$, and we have a set of $N$ samples of that function $y_i$ at locations $x_i$ where $i\in(0,1,...,N-1)$. Note that $x$ is a location in a potentially multi-dimensional domain, such as cosmological parameter space.

If the covariance between $y_i$ and $y_j$ depends only on the values of $x_i$ and $x_j$, then the distribution from which the samples were drawn is referred to as a Gaussian process. Notationally, one writes this as $y(x)={\cal GP}(\mu(x),\Sigma(x))$, where $\mu$ is the "mean function" and $\Sigma$ is the covariance matrix. Constructing a Gaussian process for regression amounts to modeling the covariance matrix using the existing samples $y_i(x_i)$, or the training data. We do not model the mean function, since the mean of the training data can always be subtracted from the training data to force $\mu=0$, and then added back on later when performing regression.

Once a covariance matrix is obtained, the predicted value of $y^*(x^*)$ comes from evaluating the conditional multivariate normal distribution given by
\begin{eqnarray}
	\label{eq:gp_regression}
    \langle y^* \rangle &=& \mu + \Sigma(x^*,\vec{x})[\Sigma (\vec{x}, \vec{x})]^{-1}\vec{y}\,, \\
    \label{eq:gp_regression2}
    {\rm Var}(y^*) &=& \Sigma(x^*,x^*) - \Sigma(x^*,\vec{x})^T[\Sigma(\vec{x}, \vec{x})]^{-1}\Sigma(\vec{x},x^*)\,.
\end{eqnarray}
In the above equation, $\vec{x}=(x_0,x_1,...,x_{N-1})$ and $\vec{y}=(y_0,y_1,...,y_{N-1})$, while $\Sigma(x_i,x_j)$ denotes the covariance between two samples $y_i$ and $y_j$ at locations $x_i$ and $x_j$. $\Sigma(x^*,\vec{x})$ is the covariance between the training data and the predicted value, and $\Sigma(x^*,x^*)$ is the covariance between points with the same input.


\subsection{Fitting function Gaussian processes}
\label{sec:fitting_function_gps}

In this work, we model the covariance matrix for the Gaussian process of each fitting function parameter as 
\begin{equation}
	\label{eq:gp_covariance}
    \Sigma = {\bf K} + {\bf I}\sigma^2_{p'}\,,
\end{equation}
where ${\bf K}$ is the kernel matrix and the second term is a diagonal matrix containing the uncertainties of the parameter under consideration $p'\in(B_0', c_0', A_1', B_1')$ from each simulation. The kernel matrix contains the covariance between two training data points computed using a kernel function. Various kernel functions exist for data with different assumed distributions. In our case, the only prior knowledge we have is that simulations that are very near to each other in cosmological parameter space should have similar sets of parameters. For this reason, a natural choice for the kernel function is the squared-exponential kernel given by
\begin{equation}
	\label{eq:squared_exponential_kernel}
    k(x,x') = k_0\exp\left[-\sum_{i=1}^{N_c} \frac{(x_i-x_i')^2}{2L_i} \right]\,.
\end{equation}
In this equation, two points in $N_c$-dimensional cosmological parameter space are specified by $x$ and $x'$, with $i$ indexing one of the $N_c$ cosmological parameters. The hyperparameters that govern the kernel are the kernel amplitude $k_0$, and each of the $N_c$ length scales $L_i$. We found that fixing $k_0=1$ to be an optimal choice in that it reduced the residual difference between the emulator predictions and the testing simulations (see \autoref{sec:validation_on_test_sims}). Each of the $L_i$ were allowed to vary, and their values were found by maximizing the likelihood
\begin{equation}
	\label{eq:gp_likelihood}
    \ln\mathcal{L} \propto -\frac{1}{2}\left[\Delta {p'}^T \Sigma^{-1}\Delta p' + \ln\det \Sigma\right]\,,
\end{equation}
where $\Delta p' = p' - \langle p' | \vec{\Omega} \rangle$ is the difference between the rotated fitting function parameter in the simulation and the predicted parameter from the Gaussian process at the cosmology of the simulation $\vec{\Omega}$ given by \autoref{eq:gp_regression}.

The procedure to reconstruct the halo bias for a given cosmology is straightforward. We use the Gaussian processes described above to predict each of the parameters $p'\in(B_0', c_0', A_1', B_1')$. Then, we take the rotation matrix described in \autoref{sec:fitting_the_simulations} to transform back to $p\in(B_0, c_0, A_1, B_1)$, use \textsc{CLASS} to map halo mass $M$ onto RMS mass variance $\sigma^2$, and then combine the fitting function parameters with \autoref{eq:tinker2010_fitting_function} and \autoref{eq:parameter_scalefactor_dependence} to predict the halo bias at any mass and redshift. At this point, the emulator is built, and we can measure its mass and redshift-dependent accuracy. 

Emulator parameters for specific cosmologies appear in \autoref{app:bias_parameters_common_cosmos}. Additionally, we provide parameters for the halo mass function emulator from \citet{McClintock2018_HMF} in \autoref{app:hmf_parameters_common_cosmos}.


\section{Emulator accuracy}
\label{sec:emulator_accuracy}

An attractive feature of Gaussian processes are their ability to estimate the uncertainty on their prediction when performing regression as given by \autoref{eq:gp_regression2}. We find that propagating this uncertainty forward through our fitting function (\autoref{eq:tinker2010_fitting_function}) significantly overestimates the uncertainty on the bias. For this reason, we estimate the accuracy of the emulator {\it a posteriori} by performing a series of tests comparing the emulator predictions directly to bias measurements in our simulations.


\subsection{Leave-one-out tests}
\label{sec:leave_one_out}

In our first set of tests, the emulator was rebuilt leaving out one training simulation at a time. We repeated this for all the training simulations used to build the emulator, and compared the residual difference between the measured bias to the predicted bias according to 
\begin{equation}
	\label{eq:residual_definition}
    \chi_R = \frac{b_{\rm measured}-\langle b \rangle}{\langle b \rangle}\,,
\end{equation}
where $b_{\rm measured}$ is the bias measured in a single mass bin in a given snapshot for one simulation, and $b_{\rm emu}$ is the bin-averaged emulator prediction for that bias. $\chi_R$ has uncertainty $\sigma_{\chi_R} = \sigma_b/\langle b \rangle$, where $\sigma_b$ is the uncertainty on the measurement. The covariance between the residuals of mass bins $i$ and $j$ within a given snapshot of a simulation given by 
\begin{equation}
	\label{eq:residual_covariance}
    {\bf C}_{\chi_R;i,j} = \frac{{\bf C}_{b;i,j}}{\langle b\rangle_{i}\langle b\rangle_{j}}\,,
\end{equation}
where ${\bf C}_{b;i,j}$ is the measured covariance between the bins as described in \autoref{sec:clustering_measurements}. The inverse-variance weighted mean absolute value of the residuals across all mass bins in all snapshots in all simulations was 1.1 percent. Note that this test only gives an upper limit on the emulator accuracy, since the tests are performed with an incomplete training sample.


\subsection{Test simulations}
\label{sec:validation_on_test_sims}

\begin{figure}
	\begin{center}
	\includegraphics[width=\linewidth]{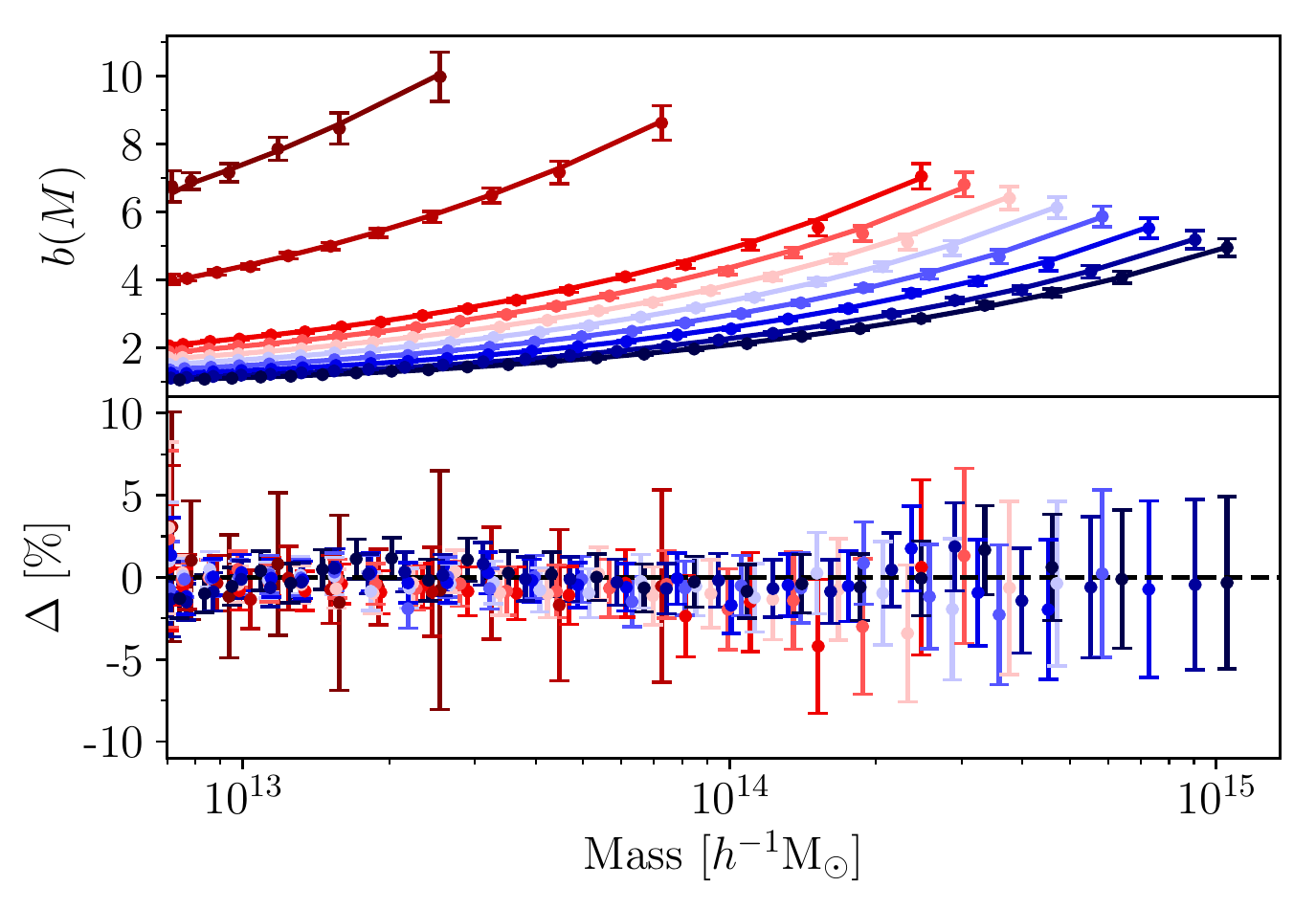}
	\caption{A comparison of the prediction of our halo bias emulator to the combined measurements in five realizations of one testing simulation cosmology. The top panel shows the direct comparison of the measured bias (points) to the emulator prediction (lines), while the bottom panel shows the fractional difference between the two. Error bars are jackknife estimates. Colors correspond to the redshift of the snapshot of a the measurement, which range from $z=3$ in dark red to $z=0$ in dark blue.
		\label{fig:testbox_comparison}}
	\end{center}
\end{figure}

As discussed in \autoref{sec:simulations}, we ran a set of test simulations at seven locations of cosmological parameter space, with five realizations each. The prediction of our emulator of the halo bias compared to the combined measurements of one location in cosmological parameter space is shown in \autoref{fig:testbox_comparison}. The inverse-variance weighted mean absolute value of the residuals across all mass bins in all snapshots in all testing simulations was 0.66 percent. This is lower than the value obtained in the leave-one-out tests due to the use of a complete emulator and reduced sample variance in the combined realizations of the test simulations.

\begin{figure}
	\begin{center}
	\includegraphics[width=\linewidth]{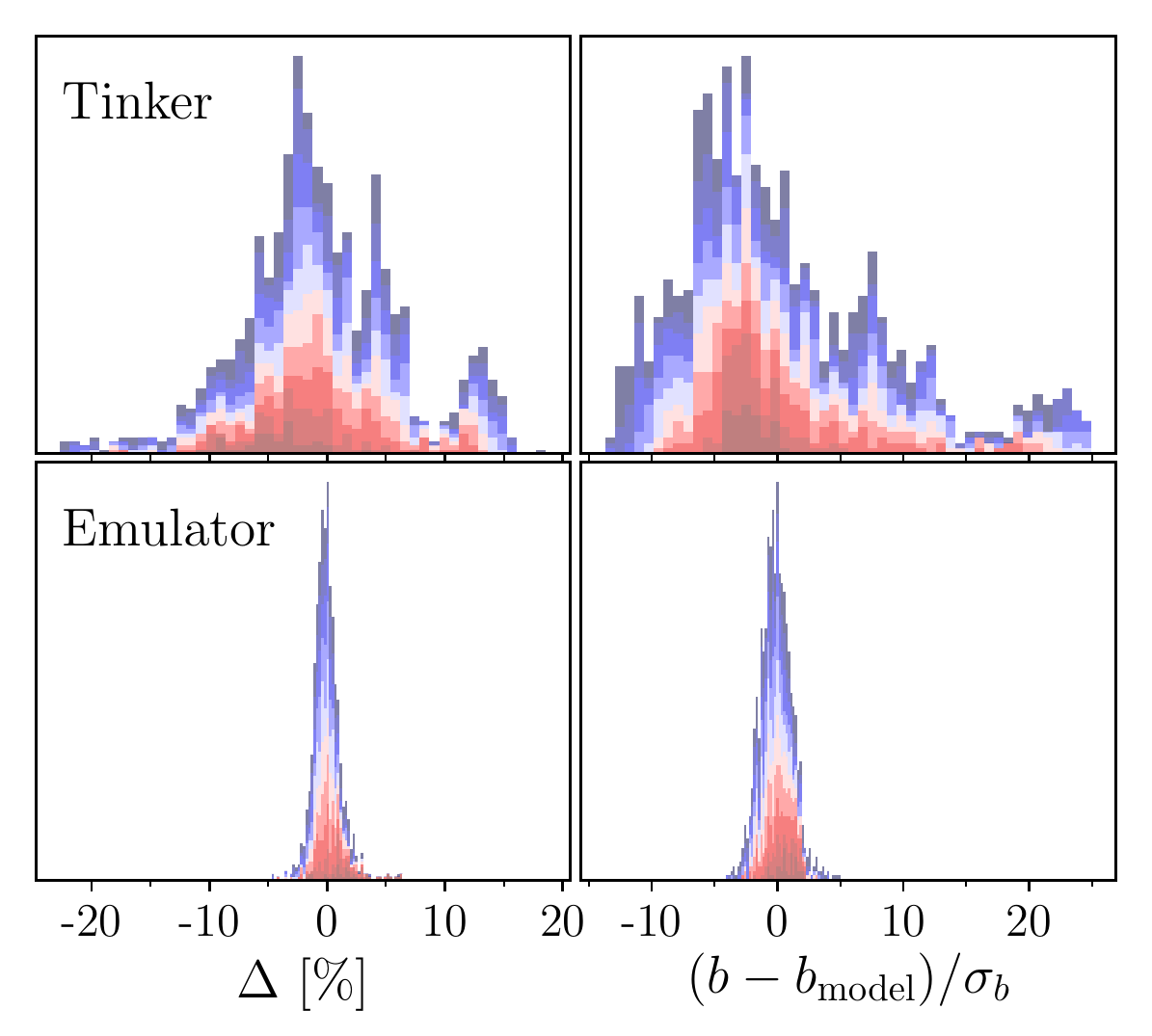}
	\caption{Residual difference between the halo bias measured in the simulation and the predictions of the \citet{Tinker2010} model in top panels, and the emulator presented in this work in the lower panels. The left panels show the percent difference between the models and simulations, while the right panels show the distribution of $\chi = \Delta b/\sigma_b$, which qualitatively reduces the effect of points with large uncertainties. Colors correspond to the redshifts of the snapshots of the measurements, with dark red corresponding to $z=3$ and dark blue to $z=0$. The ``spur'' in the residuals from the \citet{Tinker2010} model are from a simulation far from the simulations used to calibrate that model in cosmological parameter space. The emulator outperforms the \citet{Tinker2010} model at all redshifts over the mass range used to construct our emulator.
	\label{fig:bias_four_panel}}
	\end{center}
\end{figure}

We also compared the testing simulations to the predictions from the \citet{Tinker2010} model. The distribution of residuals from the emulator and the \citet{Tinker2010} model are shown in \autoref{fig:bias_four_panel}. Colors correspond to the redshifts of the snapshots. The emulator provides more accurate predictions than the \citet{Tinker2010} model. That model exibits deviations in excess of 10 percent, larger than its quoted accuracy. This stems from the simulations used in that work, which were run in a smaller cosmological parameter space, had fewer particles, and less volume. For example, the ``spur'' in the residuals for that model around 15 percent occur for a single simulation located far away in cosmological parameter space from the simulations used to calibrated the model in \citet{Tinker2010}.

\subsection{Building an accuracy model}
\label{sec:building_an_accuracy_model}

Cosmological analyses using galaxy cluster masses derived from the halo bias must propagate modeling uncertainty into the final error budget. This requirement was demonstrated in \citet{McClintock2018_HMF}, where we presented a method to account for the uncertainty in our halo mass function emulator. Even though the emulator achieves high accuracy, its uncertainty is still non-negligible. To account for this, we follow the same approach outlined in \citet{McClintock2018_HMF} and construct a generative model of the emulator accuracy conditioned on the residuals from the test simulations given by \autoref{eq:residual_definition}.

The residuals are well described by a Gaussian with zero mean and a variance that depends only on redshift. Residuals are correlated across both redshift and mass (peak height). In other words, we found no trend of accuracy of the emulator with mass or peak height, but did observe a trend with redshift. The emulator does not achieve the same level of accuracy across all redshifts for two reasons: (1) the number of halos varies across snapshots, resulting in  larger uncertinties as redshift increases, and (2) the training data is most constraining at the pivot redshift from \autoref{eq:parameter_scalefactor_dependence}. 

\begin{figure}
	\begin{center}
	\includegraphics[width=\linewidth]{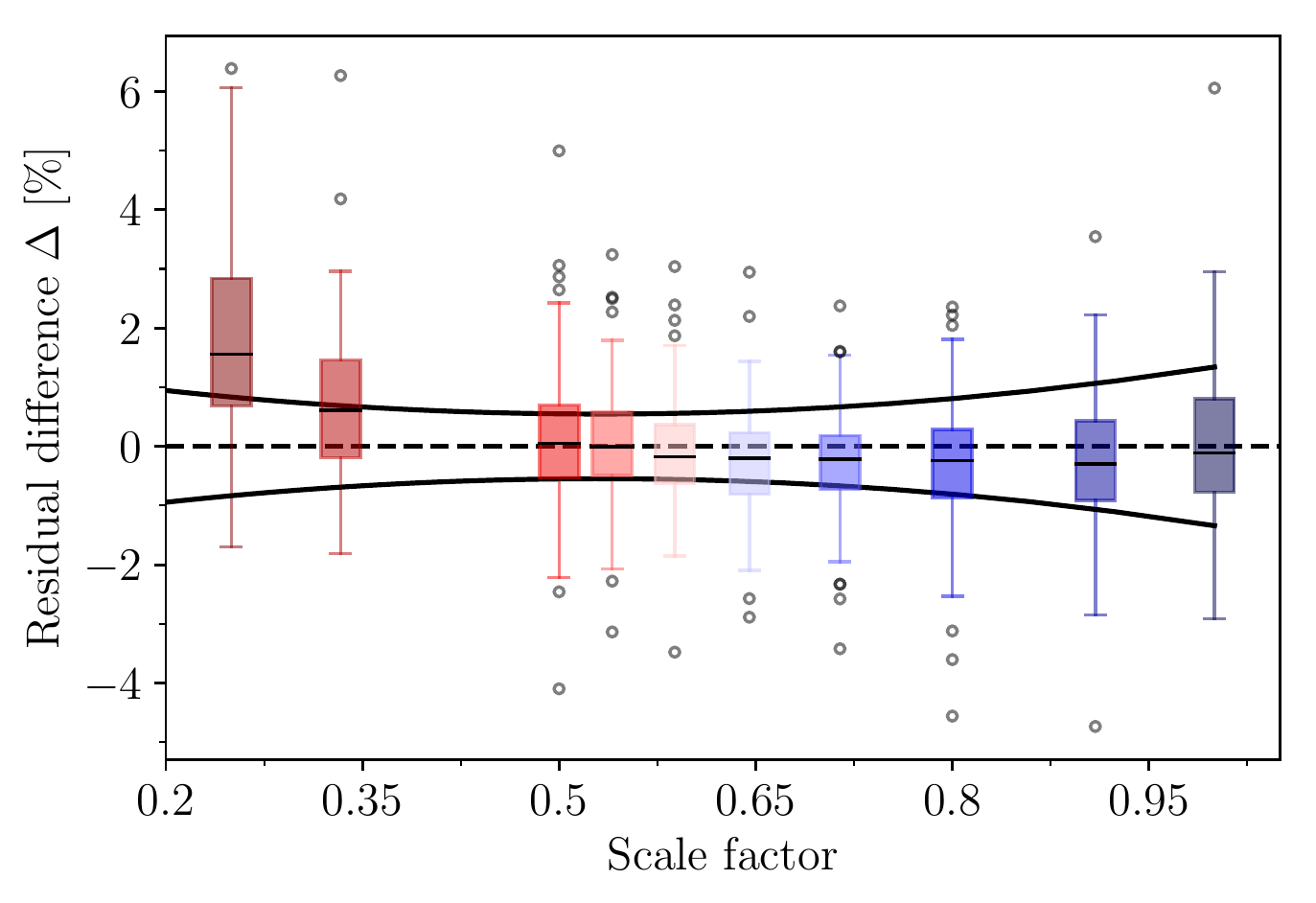}
	\caption{Model for the accuracy of the emulator, as described in \autoref{eq:residual_model} (black lines). Boxes show the middle quartiles of the residual differences between the emulator prediction and simulation results given by \autoref{eq:residual_definition}, colored by their redshift with the center line showing the median. Whiskers extend over all data except outliers, shown as points, which occur at high masses due to shot noise. The emulator accuracy did not have a mass dependence, and is most accurate at the pivot scale factor in \autoref{eq:parameter_scalefactor_dependence}. The emulator appears slightly biased at $z=3$, however the lack of halos means these residuals all have large uncertainty, which is not conveyed by the box plot.
		\label{fig:accuracy_model}}
	\end{center}
\end{figure}

The full covariance of the residuals is given as a sum of our model for the emulator accuracy and the measured covariance of the data from \autoref{eq:residual_covariance} according to
\begin{equation}
	\label{eq:residual_total_covariance}
    \tilde{\bf C}_{\rm R} = {\bf C}_{\chi_R} + {\bf C}_{\rm model}\,.
\end{equation}
Covariance between bins in different snapshots has no estimate from the simulations, meaning it is only given by ${\bf C}_{\rm model}$. The emulator accuracy and covariance are given by:
\begin{align}
	\label{eq:residual_model}
    \sigma_{\rm model}(z) &= D + E(a-a_{\rm pivot}) + F(a_{\rm p}-a_{\rm pivot})^2\,,\\
	\label{eq:residual_model2}
    {\bf C}_{\rm model}(\nu_i, \nu_j, z_i,z_j) &= e^{-|\nu_i-\nu_j|/G - |z_i-z_j|/H}\times \nonumber\\
    &\ \ \ \sigma_{\rm model}^2(z_i)\sigma_{\rm model}^2(z_j)
\end{align}
where $a_{\rm pivot} = 0.5$ is the pivot scale factor scale factor. \autoref{eq:residual_model} is also understood to be the percent accuracy of the emulator. The free parameters in the model are the accuracy at the pivot redshift $D$, the linear and quadratic coefficients of the uncertainty with scale factors $E$ and $F$, the correlation length in peak height $G$, and the correlation length in redshift $H$. These parameters are found by maximizing the likelihood
\begin{equation}
	\label{eq:residual_likelihood}
    \ln \mathcal{L}_{R} \propto -\frac{1}{2}\sum_i\left[{\bf R}^T_i \tilde{\bf C}^{-1}_{{\rm R}, i} {\bf R}_i +\ln \det\tilde{\bf C}_{{\rm R}, i} \right]\,.
\end{equation}
In this equation the sum runs over all simulations, ${\bf R}_i$ is a vector containing all residuals in the $i$-th testing simulation across all snapshots, and the covariance matrix represents the covariance between all residuals in that simulation. The resulting model is shown in \autoref{fig:accuracy_model}, with box plots showing the distribution of the residuals at each scale factor. The emulator accuracy is better than 1 percent for $z\in[0.1,1]$. This is the relevant redshift range for cluster cosmology in DES \citep{Melchior2017,McClintock2018,Costanzi2018_DESSDSSClusterCosmo}, and is sufficient for performing a 2 percent calibration of halo masses as seen in \autoref{fig:bias_accuracy}.

Using the model for the residuals, we can make random realizations of the residuals of the bias. This allows for the uncertainty in the bias emulator to be properly propagated when using the bias emulator, allowing for accurate estimates of error budgets. We provide the means to create residuals along with the bias emulator. In order to make these realizations, we can draw from a Gaussian with zero mean and covariance given by ${\bf C}_{\rm model}$. Note that ${\bf C}_{\chi_R}$ is only used for optimizing the residual model, and is not involved in making realizations of the uncertainty.


\section{Building models with the halo bias}
\label{sec:using_the_emulator}

Our emulator can be combined with an estimate of the matter correlation function $\ximm$ in order to model the spatial distribution of halos such as the halo--matter or halo--halo correlation functions. One way to estimate the matter correlation function is to predict the matter power spectrum $P(k,z)$ and perform a Fourier transform according to
\begin{equation}
	\label{eq:matter_correlation_function_definition}
    \xi_{\rm mm}(r,z) = \int_0^{\infty} \frac{{\rm d}k}{k} \frac{k^3P(k,z)}{2\pi^2}j_0(kr)\,,
\end{equation}
where $j_0(kr)$ is 0th spherical Bessel function of the first kind. In the absence of nonlinear growth of structure, the matter power spectrum is given by the linear matter power spectrum $P_{\rm lin}$ as computed by, e.g., CAMB or CLASS \citep{CAMB,CLASS1}. As structure forms, the linear matter power spectrum is no longer an accurate representation of the true power spectrum. Instead, the true power spectrum closely resembles $P_{\rm nl}$, which includes modifications from the gravitational nonlinearities \citep{Smith2003_halofit,Takahashi2012_halofit,Smith2018_NGenHalofit}. 
We combined our bias emulator with the Fourier transformed prediction of the nonlinear power spectrum from Halofit, as implemented in CLASS. In doing so, we successfully model $\xihm$ at scales $\sim$20 $\hmpc$ and above. Below this scale, we require an emulator for $\ximm$, which we will present in a future work.

\begin{figure}
	\begin{center}
	\includegraphics[width=\linewidth]{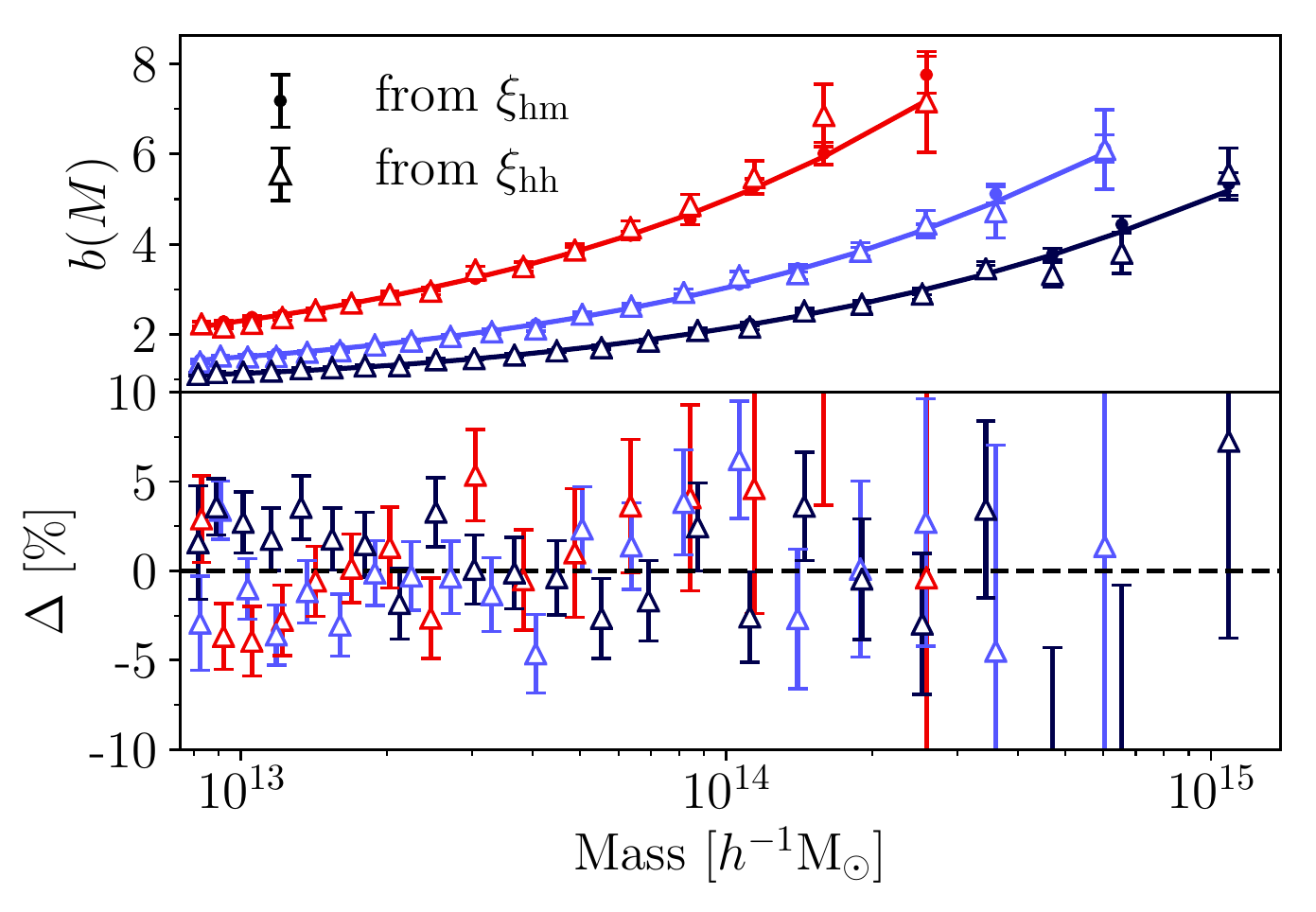}
	\caption{Halo bias computed from the ratio of the halo--halo correlation function $\xihh$ or the halo--matter correlation function $\xihm$ to the matter--matter correlation function $\ximm$. The top panel plots the bias for the same mass bins at three redshifts $z\in[1.0,0.4,0]$, where red(blue) is high(low) redshift. The line shows the predicted halo bias as a function of redshift at each mass from our emulator. The lower panel shows the percent difference between the bias measured from $\xihh$ and the emulator prediction. The emulator prediction and bias from $\xihh$ are consistent, making our emulator suitable for cluster clustering.
		\label{fig:hh_vs_hm}}
	\end{center}
\end{figure}

We also tested that our bias emulator is suitable for modeling halo--halo clustering by measuring 
\begin{equation}
	\label{eq:halo_autocorrelation}
    b^{2}(M) = \frac{\xihh}{\ximm}\bigg|_{\rm 10-40\ {\it h}^{-1}Mpc}
\end{equation}
where $\xihh$ is the halo--halo correlation function. When computing $b^2$ we use the same radial scales as in \autoref{eq:halo_definition_xi}. \autoref{fig:hh_vs_hm} shows the bias computed using both $\xihm$ and $\xihh$ within the same mass bins at three redshifts within a testing simulation. The lines show the emulator prediction as a function of mass for each snapshot. The filled points show the bias measured from $\xihm$, while the unfilled points show the bias measured from $\xihh$. We use the same mass bins for both sets of measurements. The predictions and measurements all agree, meaning our halo bias emulator is suitable for use in cluster clustering analyses.

\section{Testing higher and lower masses}
\label{sec:comparison_to_other_simulations}

We ran simulations with more volume and higher resolution than our training set in order to assess emulator performance at higher and lower masses. At low mass, we tested the emulator against our high-resolution simulations. These are 400 $\hmpc$ per side with 2048$^3$ particles, and resolve halos down to $\sim$10$^{11}\ \hmsun$. We also ran simulations with 3 $\hgpc$ per side and 2048$^3$ particles to acquire improved statistics for high mass halos. The cosmological parameters for the high resolution simulations are different than the training and testing simulations, while the large volume simulations have the same parameters as the testing simulations. Neither set of simulations were used to train the emulator.

\begin{figure*}
	\begin{center}
	\includegraphics[width=\linewidth]{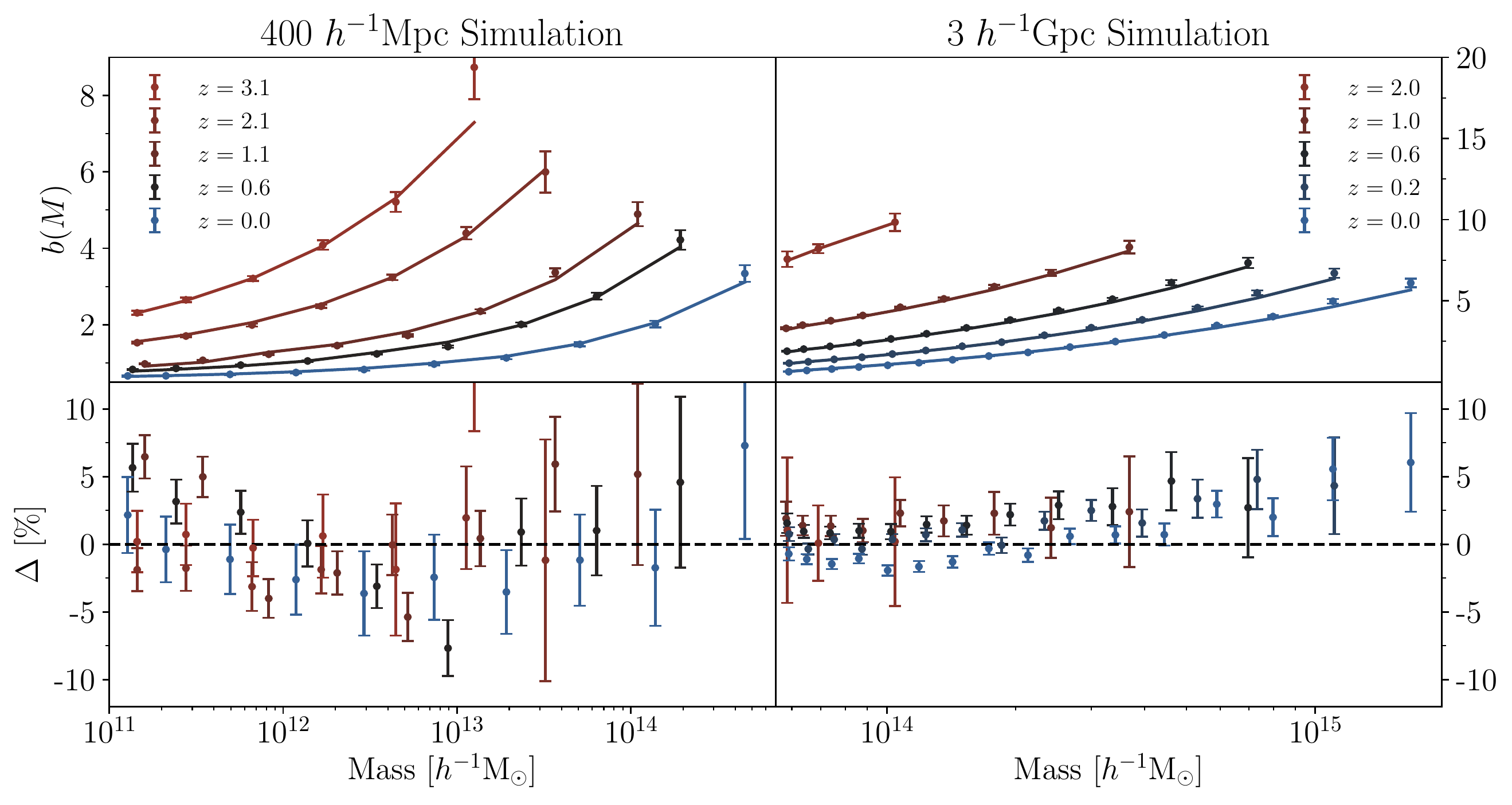}
	\caption{Emulator predictions compared to the bias measured in a $(400\ \hmpc)^3$ high resolution (left) and a $(3\ \hgpc)^3$ large volume (right) simulation. The high resolution simulation resolves halos down to $\sim$10$^{11}\ \hmsun$, meaning the emulator extrapolates past its training data by two orders of magnitude. For masses in peaks below the minimum peak height in our training data, the bias emulator asymptotes to the \citet{Tinker2010} model (see \autoref{sec:comparison_to_other_simulations}). The large volume resolved many more halos, making measurements of the bias very precise. These simulations are well described by our emulator, despite having significantly better statistics than any individual training or testing simulations. This is demonstrated by all simulations having acceptable $\chi^2$ per degree of freedom when accounting for the emulator accuracy.
		\label{fig:other_sims_comaparison}}
	\end{center}
\end{figure*}

We measured the linear halo bias in both sets of simulations in the same manner as our training and testing simulations as described in \autoref{sec:clustering_measurements}. For both sets simulations, we used 1000 spatial subregions for jackknifing. \autoref{fig:other_sims_comaparison} shows measurements of the halo bias in one simulation from each set, as well as the prediction from the emulator.

The high-resolution simulations extend roughly two orders of magnitude lower in mass compared to our training simulations. Our emulator originally was not able to extrapolate accurately down to low masses at low redshifts. To address this, we employ an extrapolation scheme that allows the prediction to asymptote to the \citet{Tinker2010} model at very low peak heights. The simulations used to construct the halo bias model presented in \citet{Tinker2010} resolved halos down to $10^{10}\ \hmsun$, making it better suited for this regime, despite the scatter in the model shown in \autoref{fig:bias_four_panel}. At high redshifts the emulator extrapolated well. These two features: poor performance at low mass at low redshift and exceptional performance at high mass, indicates that the emulator performs well in the highest peaks of the density field, regardless of the mass of the halo that resides in that peak.

In detail, our low mass extrapolation scheme is as follows:
\begin{enumerate}
    \item Given $\Omega_m$, calculate the minimum halo mass we would have resolved in our training simulations according to \autoref{eq:particle_mass}.
    \item Calculate the minimum peak height for this mass given this cosmological model according to $\nu_{\rm min} = \nu(M,z=0)=\delta_c/\sigma(M,z=0)$, using \autoref{eq:sigma_definition}. At fixed mass, the minimum peak height will be at $z=0$.
    \item Calculate the peak height at the halo mass and redshift we are interested in from $\nu(M,z)=\delta_c/\sigma(M,z)$.
    \item If $\nu(M,z) < \nu_{\rm min}$, then extrapolate the halo bias according to
    \begin{equation}
        \label{eq:bias_extrapolation}
        b(\nu,z) = b_{\rm emulator}(\nu_{\rm min}, z) - b_{\rm Tinker}(\nu_{\rm min},z) + b_{\rm Tinker}(\nu, z)\,.
    \end{equation}
\end{enumerate}
This behaviour is seen in \autoref{fig:other_sims_comaparison} in the left panels, where the prediction at low mass ($\sim 10^{10}-10^{12}\ \hmsun$) is from the emulator at high redshift and from the \citet{Tinker2010} model at low redshifts.
Without this extrapolation scheme, we find 20\% residuals at low redshift due to the inability of the emulator to extrapolate to very low peak heights.
We plan to update our emulator to cover a larger mass range when our high resolution simulation suite is complete. For now, this approach allows our augmented emulator to be used for all mass scales probed by the high resolution simulations.

Our large volume simulations reduce sample variance at high mass by a factor of $\sim$27. For this reason the residuals in the lower right panel of \autoref{fig:other_sims_comaparison} are $\sim$5 times smaller than those in the training simulations. While the measured biases appear to be systematically offset from our emulator prediction, the data are fully consistent with our model when we account for the modeling accuracy. We compute $\chi^2$ of these simulations according to
\begin{equation}
    \label{eq:large_volume_chi2}
    \chi^2 = {\bf R}^T\tilde{\bf C}_{\rm R}^{-1}{\bf R}\,,
\end{equation}
where ${\bf R}$ is a vector of the residuals and $\tilde{\bf C}_{\rm R}$ is the covariance matrix of the residuals according to \autoref{eq:residual_total_covariance}. When including the accuracy model in the total covariance, the $\chi^2$ per degree of freedom of is acceptable for all snapshots in all simulations. We checked that the $\chi^2$ of each simulation accounting for correlations between snapshots and found these was also acceptable. Therefore, while simulations with larger volume can be used to measure the halo bias more precisely than our training simulations, the bias in our $3\ \hgpc$ is consistent with the emulator predictions up to its accuracy limits.

\section{Conclusions}
\label{sec:conclusions}

In this work we present an emulator for the linear halo bias, $b(M,z)$. The emulator is trained on a large suite of $N$-body simulations spanning a volume in cosmological parameter space given by the 3$\sigma$ contours Planck+WMAP+BAO+SNIa. The training simulations resolve halos down to $10^{13}\ \hmsun$, and using an extrapolation scheme whereby the emulator asymptotes to the \citet{Tinker2010}, it is able to accurately predict the halo bias down to $10^{11}\ \hmsun$. We present a model for the emulator accuracy, and detail the steps required to propagate the uncertainty in our model forward in a larger analysis. This accuracy is evaluated by a set of test simulations with a factor of five less sample variance than our training simulations, as well as simulations with more particles that probe more volume and higher resolution than our fiducial set. In all cases, the emulator is able to accurately predict the halo bias at all points in cosmological parameter space. We demonstrate the ability to model the spatial distribution of halos as described by the halo--matter and halo--halo correlation functions.

This emulator is useful for calibrating the mass of galaxy clusters or large galaxies, for instance in weak lensing or clustering analyses \citep{Baxter2016, Jimeno2017_RMCC, Melchior2017, McClintock2018}. Before this work, theoretical uncertainty in linear halo bias models were a limiting factor for some experiments \citep[e.g.][]{TinkerM2N2012, Baxter2016}. Our emulator achieves sub-percent accuracy for all redshifts except at $z=0$, where it is only slightly above as seen in \autoref{fig:accuracy_model}. This is far below the level of other systematics that contribute to current cluster mass calibrations. If these systematics can be controlled for, our emulator can be used to perform a 2 percent measurement of cluster masses, as shown in \autoref{fig:bias_accuracy}.

This paper is part of a series that presents stand-alone emulators from the \aemulus\ Project. The clustering measurements made here will be used in the future for constructing emulators for the halo--matter, halo--halo, and matter--matter correlation functions. When complete, this suite of tools will be useful for many applications outside of cluster cosmology, such as detecting halo splashback boundaries \citep{DiemerSplashback2014, Chang2018_splashback, Shin2018_SZSplashback} and covariance matrix estimation \citep{Gruen2015, Krause2017_DESCovs}. At that point, we will have enabled accurate and precise modeling of cluster observables, and drastically reduced theoretical modeling uncertainty. This is an essential step toward precision galaxy cluster cosmology in present and future experiments.

\acknowledgments
ER is supported by DOE grant DE-SC0015975 and the Sloan Foundation grant FG-2016-6443. JD, RHW, SM, MRB received support from the U.S. Department of Energy under contract number DE-AC02-76SF00515. This research used resources of the National Energy Research Scientific Computing Center, a DOE Office of Science User Facility supported by the Office of Science of the U.S. Department of Energy under Contract No. DE-AC02-05CH11231. ZZ acknowledges support by NASA grant 15-WFIRST15-0008 Cosmology with the High Latitude Survey WFIRST Science Investigation Team (SIT).

This research used resources of the National Energy Research Scientific Computing Center, a DOE Office of Science User Facility supported by the Office of Science of the U.S. Department of Energy under Contract No. DE-AC02-05CH11231. Some of the computing for this project was performed on the Sherlock cluster. We would like to thank Stanford University and the Stanford Research Computing Center for providing computational resources and support that contributed to these research results. This research made use of computational resources at SLAC National Accelerator Laboratory, and the authors thank the SLAC computational team for support.


\software{Python,
Matplotlib \citep{matplotlib},
NumPy \citep{numpy},
SciPy \citep{scipy},
emcee \citep{Foreman13},
george \citep{Ambikasaran2014},
CAMB \citep{CAMB},
CLASS \citep{CLASS1},
\textsc{Rockstar} \citep{Behroozi2013},
GADGET \citep{Springel05},
2LPT \citep{Crocce2006}
}

\bibliographystyle{yahapj}
\bibliography{astroref,software}

\appendix


\section{Required bias emulator accuracy}
\label{app:required_bias_emulator_accuracy}

The required accuracy on models of the halo bias is set by the precision with which we can measure halo masses in real data. Assuming that halo masses can be measured with a precision $\sigma_{\ln M} = \sigma_M/M$, the corresponding uncertainty in the halo bias is
\begin{equation}
    \label{eq:halo_bias_required_accuracy}
    \sigma_{\ln b} = \frac{M}{b}\frac{{\rm d}b}{{\rm d}M} \sigma_{\ln M}\,.
\end{equation}
If all information about the halo mass came exclusively from the two-halo term of the appropriate halo statistics, the above estimate of the uncertainty corresponds to the calibration requirement for the halo bias function. That is, the above equation provides a conservative estimate of the necessary precision of the bias emulator given a target precision for mass calibration.  The precision estimate is both mass and redshift dependent, with lower-mass halos requiring a higher precision calibration. Given the fixed precision of the bias emulator, we can readily compute the minimum mass $M_{\rm min}$ above which our emulator is sufficiently precise given expected uncertainties in mass calibration.

In \autoref{fig:bias_accuracy} we compute the minimum halo mass $M_{\rm min}$ for which our emulator precision is sufficiently high for varying levels of precision in mass calibration, namely $\sigma_{\ln M}\in[0.015, 0.02, 0.03]$. Additionally, we plot the minimum halo mass in our training simulations as the black dashed line. We can see that the emulator is easily capable of modeling the halo bias for modern cluster analyses ($M\gtrsim 10^{14}\ M_{\odot}$) until mass calibration uncertainties reach the $\approx 2\%$ level.  Because our calculation is conservative, we expect our bias emulator will continue to suffice even after that, but more detailed calculations will need to be carried out at that point.


\section{Bias emulator parameters for common cosmologies}
\label{app:bias_parameters_common_cosmos}

Here we provide parameter predictions from the emulator for sets of cosmological parameters that are frequently used. This is meant for users that require an accurate model of halo bias at only a single cosmology. Listed below in \autoref{tab:emulator_parameters_cosmos} are only the parameters that are predicted by the emulator. The fixed parameters in the halo bias are listed in the main text in \autoref{tab:parameter_values}. The cosmologies considered are shown in \autoref{tab:cosmo_parameters} and are presented in
\citet{Planck2015_CosmologicalParams, Planck2018_CosmologicalParams} and \citet{DESY1_KP}. JLA is the ``Joint Light-curve Analysis'' of supernovae described in \citet{JLA2014_CosmoParams}. BAO designates baryon acoustic oscillation analyses performed in \citet{Anderson2014}, \citet{Beutler2011_BAO} and \citet{Ross2015_BAO}. The $H_0$ constraint comes from distance latter analysis performed in \citet{Riess2011_Ladder}. For DES, the $w$CDM result comes from a joint analysis with the \citet{Planck2015_CosmologicalParams} data. All Planck results refer to the marginalized means of the TT,TE,EE+lowE+lensing analyses, however CMB lensing is removed when combining with DES. Note that constraints on $w$ and $N_{\rm eff}$ are made independently in all analyses and are not varied together, but this results in only a small change in the emulator prediction compared to fixing $N_{\rm eff}$ to its theoretically predicted value of $3.046$.

\begin{table}[ht]
    \centering
    \setlength{\tabcolsep}{.4em}
	\caption{Cosmological parameters for selected experiments for which we provide bias emulator parameters in \autoref{tab:emulator_parameters_cosmos} and mass function emulator parameters in \autoref{tab:emulator_parameters_cosmos_MF}.
	\label{tab:cosmo_parameters}}
	\begin{tabular}{l|llllllll}
	Parameters & $\Omega_b h^2$ & $\Omega_{c}h^2$ & $w$ & $n_s$ & $\ln(10^{10}A_s)$ & $H_0$ & $N_{\rm eff}$\\ \hline
    Planck (2015) $\Lambda$CDM & 0.022 & 0.118 & -1 & 0.968 & 3.064 & 67.9 & 3.150 \\
	Planck (2015)+JLA+BAO+$H_0$ $w$CDM & 0.022 & 0.118 & -1.006 & 0.968 & 3.064 & 67.9 & 3.150 \\
	Planck (2018) $\Lambda$CDM & 0.022 & 0.120 & -1 & 0.965 & 3.044 & 67.36 & 2.990 \\
	Planck (2018)+JLA+BAO $w$CDM & 0.022 & 0.120 & -1.028 & 0.965 & 3.044 & 67.36 & 2.990 \\
	DES+JLA+BAO $\Lambda$CDM & 0.023 & 0.116 & -1 & 1.050 & 3.043 & 67.9 & 3.046 \\
	DES+Planck+JLA+BAO $w$CDM & 0.022 & 0.117 & -1.00 & 0.973 & 3.186 & 67.36 & 3.046 \\
	\end{tabular}
\end{table}

\begin{table}[ht]
	\centering
	\setlength{\tabcolsep}{.4em}
	\caption{Bias emulator parameters given the cosmological parameters found in specific analyses.
	\label{tab:emulator_parameters_cosmos}}
	\begin{tabular}{l|llll}
	    Analysis & $B_0$ & $c_0$ & $A_1$ & $B_1$\\ \hline
	Planck (2015) $\Lambda$CDM & 1.147 & -1.192 & 0.282 & 0.748 \\
	Planck (2015)+JLA+BAO+$H_0$ $w$CDM & 1.145 & -1.189 & 0.272 & 0.743 \\
	Planck (2018) $\Lambda$CDM & 1.132 & -1.071 & 0.230 & 0.746 \\
	Planck (2018)+JLA+BAO $w$CDM & 1.122 & -1.044 & 0.179 & 0.727 \\
	DES+JLA+BAO $\Lambda$CDM & 1.166 & -1.400 & 0.264 & 0.726 \\
	DES+Planck+JLA+BAO $w$CDM & 1.113 & -0.938 & 0.400 & 0.802 \\
	\end{tabular}
\end{table}

\section{Mass function emulator parameters for common cosmologies}
\label{app:hmf_parameters_common_cosmos}

Here we provide emulator parameters for the mass function emulator described in \citet{McClintock2018_HMF} at the cosmologies listed above in \autoref{app:bias_parameters_common_cosmos}. The fixed parameters in that emulator are $d_0=2.393$ and $f_1=0.116$. See that paper for implementation details.

\begin{table}[ht]
	\centering
	\setlength{\tabcolsep}{.4em}
	\caption{Mass function emulator parameters given the cosmological parameters found in specific analyses.
	\label{tab:emulator_parameters_cosmos_MF}}
	\begin{tabular}{l|llllll}
	    Analysis & $e_0$ & $f_0$ & $g_0$ & $d_1$ & $e_1$ & $g_1$\\ \hline
	Planck (2015) $\Lambda$CDM & 0.865 & 0.502 & 1.253 & 0.239 & 0.170 & 0.113 \\
	Planck (2015)+JLA+BAO+$H_0$ $w$CDM & 0.866 & 0.501 & 1.252 & 0.229 & 0.167 & 0.106 \\
	Planck (2018) $\Lambda$CDM & 0.864 & 0.503 & 1.254 & 0.268 & 0.175 & 0.121 \\
	Planck (2018)+JLA+BAO $w$CDM & 0.867 & 0.500 & 1.246 & 0.226 & 0.164 & 0.090 \\
	DES+JLA+BAO $\Lambda$CDM & 0.862 & 0.513 & 1.260 & 0.198 & 0.176 & 0.108 \\
	DES+Planck+JLA+BAO $w$CDM & 0.899 & 0.495 & 1.263 & 0.209 & 0.192 & 0.141 \\
	\end{tabular}
\end{table}

\end{document}